\documentclass[twocolumn,aps,prl,amsmath,amssymb]{revtex4-1}

\usepackage{graphicx}
\usepackage{dcolumn}
\usepackage{bm}
\usepackage{amsmath}
\usepackage{amsfonts}
\usepackage[usenames,dvipsnames,svgnames,table]{xcolor}
\DeclareGraphicsExtensions{.pdf,.eps,.png,.jpg,.mps}

\usepackage[colorlinks]{hyperref}
\hypersetup{
	pdfstartview={FitH},
	linkcolor=blue,
	citecolor=blue,
	filecolor=blue,
	urlcolor=blue
}

\begin{document}

\title{Valley and spin dynamics in monolayer MoS$_{2}$}

\author{S. Dal Conte$^{1,2}$}
\author{F. Bottegoni$^{2}$,E.A.A. Pogna$^2$}
\author{S. Ambrogio$^3$}
\author{I. Bargigia$^4$, C. D'Andrea$^{2,4}$}
\author{D. De Fazio$^5$, A. Lombardo$^5$, M. Bruna$^5$}
\author{F. Ciccacci$^2$}
\author{A.C. Ferrari$^5$}
\author{G. Cerullo$^{1,2}$}
\author{M. Finazzi$^2$}
\affiliation{$^1$IFN-CNR, Piazza L. da Vinci 32, I-20133 Milano, Italy}
\affiliation{$^2$Dipartimento di Fisica, Politecnico di Milano, Piazza L. da Vinci 32, I-20133 Milano, Italy}
\affiliation{$^3$Dipartimento di Elettronica, Informatica e Bioingegneria, Politecnico di Milano and IU.NET, I-20133 Milano, Italy}
\affiliation{$^4$Center for Nano Science and Technology@PoliMi, Istituto Italiano di Tecnologia, via Giovanni Pascoli 70/3, 20133 Milan, Italy}
\affiliation{$^5$Cambridge Graphene Centre, University of Cambridge, 9 JJ Thomson Avenue, Cambridge CB30FA, UK}

\begin{abstract}
Valleytronics targets the exploitation of the additional degrees of freedom in materials where the energy of the carriers may assume several equal minimum values (valleys) at non-equivalent points of the reciprocal space. In single layers of transition metal dichalcogenides (TMDs) the lack of inversion symmetry, combined with a large spin-orbit interaction, leads to a conduction (valence) band with different spin-polarized minima (maxima) having equal energies. This offers the opportunity to manipulate information at the level of the charge (electrons or holes), spin (up or down) and crystal momentum (valley). Any implementation of these concepts, however, needs to consider the robustness of such degrees of freedom, which are deeply intertwined. Here we address the spin and valley relaxation dynamics of both electrons and holes with a combination of ultrafast optical spectroscopy techniques, and determine the individual characteristic relaxation times of charge, spin and valley in a MoS$_{2}$ monolayer. These results lay the foundations for understanding the mechanisms of spin and valley polarization loss in two-dimensional TMDs: spin/valley polarizations survive almost two-orders of magnitude longer for holes, where spin and valley dynamics are interlocked, than for electrons, where these degrees of freedom are decoupled. This may lead to novel approaches for the integration of materials with large spin-orbit in robust spintronic/valleytronic platforms.
\end{abstract}

\maketitle
Exploiting and manipulating the spin and valley degrees of freedom of carriers in order to process and store information is one of the most challenging goals of modern solid-state physics, which has already resulted in the demonstration of several functional devices\cite{Zutic2004,Yang2008,Ilgaz2010,Demidov2012,Dery2007a,Bottegoni2014}. In this context, transition metal dichalcogenides (TMDs) add novel functionalities, due to the strong interplay between the spin and the crystal momentum of the carriers\cite{Xiao2012}, and represent a promising platform to develop new spin- and valleytronic devices thanks to their peculiar electronic structure\cite{Zutic2014} and the integrability with graphene technology\cite{Ferrari2014,Roy2013}.

\begin{figure}
\centerline{\includegraphics[width=75mm]{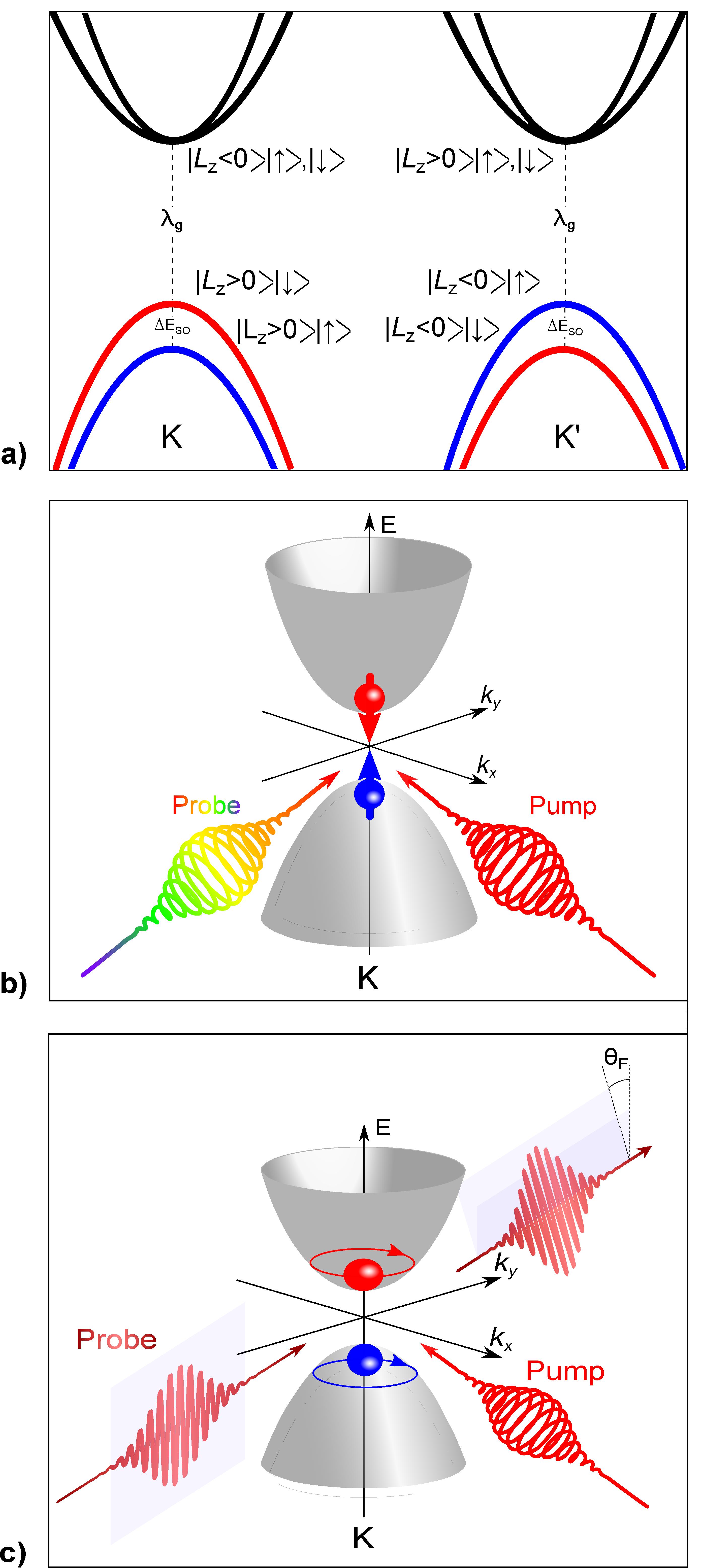}}
\caption{\label{fig1}a) 1L-MoS$_{2}$ band structure around \textbf{K} and \textbf{K'} \cite{Kosmider2013,Kormanyos2013}. The electronic states are labeled as a function of the projection of their average angular momentum along the direction perpendicular to the plane and the spin angular momentum. b), Sketch of the TRCD experiment. A circularly polarized pump pulse, resonant with the bandgap, creates a population of spin-oriented \textit{e} and \textit{h} in \textbf{K} or \textbf{K'} due to valley selectivity. Co- and counter-circularly polarized broadband pulses probe the transient variation of the transmission spectrum due to the pump-induced phase space filling effect. The TRCD dynamics is associated with spin-flip and valley-scattering processes (the red and blue thick arrows represent the \textit{e} and \textit{h} spin, respectively). c), Sketch of the TRFR experiment. A spin polarized \textit{e} and \textit{h} population is initialized by a circularly polarized pump pulse. The rotation angle $\theta\rm_{F}$ of the probe linear polarization is related to the net value of the orbital momentum projection of both carriers (indicated by the red and blue circular arrows). Since the states at \textbf{K} and \textbf{K'} have opposite orbital momentum projections, TRFR is sensitive exclusively to intervalley scattering processes.}
\end{figure}

In a single MoS$_{2}$ layer (1L-MoS$_{2}$) both the minimum of the conduction band (CB) and the maximum of the valence band (VB) are located at the \textbf{K} and \textbf{K'} points of the hexagonal Brillouin zone (Fig.\ref{fig1}a)\cite{Kosmider2013,Kormanyos2013}, allowing for direct absorption transitions in the visible range\cite{Mak2010}. The lack of inversion symmetry, combined with the C$_{3h}$ symmetry of the Bloch wavefunctions at \textbf{K, K'}, lead to electron states with a non-vanishing projection of their average angular momentum $\left\langle L_{z}\right\rangle$ along the direction perpendicular to the MoS$_{2}$ plane\cite{Kosmider2013,Kormanyos2013}. In particular, at \textbf{K}, $\left\langle L_{z}\right\rangle<$0 for states contributing to the CB minimum, while $\left\langle L_{z}\right\rangle>$0 for those at the VB maximum. At \textbf{K'} the signs are inverted, since \textbf{K, K'} are related to each other by time-reversal conjugation\cite{Kosmider2013,Kormanyos2013}. The valley index can be thus regarded as a discrete degree of freedom for low-energy carriers, robust against defects, contaminants, and low-energy phonons because of the large valley separation in momentum space\cite{Zeng2012}, in principle enabling valley-based noise-resistant quantum computation\cite{Culcer2012}.

It is important to note that $\left\langle L_{z}\right\rangle$ for states at the VB maxima is associated with the transition metal (Mo) \textit{d} orbitals\cite{Sallen2014}. The strong spin-orbit interaction acting on the latter induces a $\Delta E_{\mathrm{SO}}=$160meV energy splitting between spin-up and spin-down states, and determines opposite spin polarizations at the VB \textbf{K, K'} (Fig.\ref{fig1}a)\cite{Kosmider2013,Kormanyos2013}. Conversely, $\left\langle L_{z}\right\rangle$ at the CB minima is provided by a small contribution from the \textit{p} atomic orbitals of the chalcogen (S) atoms (see Table II in Ref.\onlinecite{Kosmider2013}), resulting in a small spin-orbit splitting (a few meV)\cite{Kosmider2013,Kormanyos2013}. States with opposite spin at the bottom of the CB band are thus almost degenerate, meaning that the spin of conduction electrons is independent of the occupied valley\cite{Kosmider2013,Kormanyos2013}. This strong coupling between momentum and spin makes single-layer TMDs particularly appealing when compared to other valleytronics platforms, such as those based on silicon\cite{Culcer2012} or graphene\cite{Rycerz2007}. The large spin-orbit interaction, combined with optical excitation by circularly polarized light, allows the photogeneration of electron (\textit{e}) and hole (\textit{h}) populations in the CB and VB, respectively, with up to 100$\%$ spin and valley polarization\cite{Mak2012,Xiao2012}.

Thus, studying the dynamics of the relaxation of valley and spin polarization for both photoexcited e/h is crucial to understand the mechanisms leading to the loss of information that could be associated with these degrees of freedom. Thanks to their peculiar electronic structure, TMDs offer the unprecedented opportunity of comparing the dynamics of VB holes, where spin and valley dynamics are interlocked\cite{Kosmider2013,Kormanyos2013}, with that of CB electrons, for which spin and valley index are completely decoupled\cite{Kosmider2013,Kormanyos2013}. This allows for the direct evaluation of the stabilization that the coupling between spin and valley provides against perturbations.

The relaxation dynamics in TMDs was previously investigated by time-resolved photoluminescence (PL)\cite{Korn2011,Lagarde2014} or pump-probe spectroscopy\cite{Wang2013b,Mai2014,Zhu2014}. However, due to the low temporal resolution in Refs.\onlinecite{Korn2011,Lagarde2014} and/or to the lack of selectivity in Refs.\onlinecite{Wang2013b,Mai2014,Zhu2014}, these studies were not able to unambiguously distinguish all the scattering (spin and valley) relaxation channels in MoS$_{2}$, and to decouple the \textit{e} and \textit{h} dynamics.

Here we determine the individual characteristic relaxation times of charge, spin and valley-associated crystal momentum in a 1L-MoS$_{2}$. We find that the \textit{h} spin/valley relaxation time is over an order of magnitude larger than for \textit{e}. This stems from the fact that \textit{e} can exchange spin and valley degrees of freedom independently, while \textit{h} have to undergo a scattering event where spin \textit{and} valley need to be \textit{simultaneously} transferred to the lattice. Our results demonstrate that coupling valley and spin in high spin-orbit materials is a promising strategy towards the realization of robust spintronic/valleytronic platforms, where information associated with spin/valley degrees of freedom can thus survive for a longer time.

To investigate the 1L-MoS$_{2}$ spin and valley dynamics, we must separately access, for both \textit{e} and \textit{h}, the spin relaxation processes occurring within the same valley and those involving intervalley scattering. To do so, we combine two complementary ultrafast optical spectroscopy techniques: Time-Resolved Circular Dichroism (TRCD)\cite{Hilton2002} (see caption of Fig.\ref{fig1}b for an outline of this technique) and Time-Resolved Faraday Rotation (TRFR)\cite{Dyakonov2008} (see caption of Fig.\ref{fig1}b for an outline of this technique). We find that \textit{e} relaxation, involving both spin-flip and inter-valley scattering, occurs on a short time ($\sim$200 fs), whereas \textit{h} inter-valley scattering acts on longer time scales$\sim$5ps. This indicates that the non-equilibrium dynamics in 1L-MoS$_{2}$, including the recombination processes, is dominated by \textit{h} scattering.

A 1L-MoS$_{2}$ flake ($\sim10\mu$m$\times30\mu$m, see Methods) is mechanically exfoliated from bulk MoS$_{2}$, identified and characterized by optical contrast, PL and Raman measurements, and then transferred on a fused silica substrate (see Methods). The use of a metal frame, as explained in Methods, allows us to locate the same flake for both TRCD and TRFR measurements.

\begin{figure}
\centerline{\includegraphics[width=70mm]{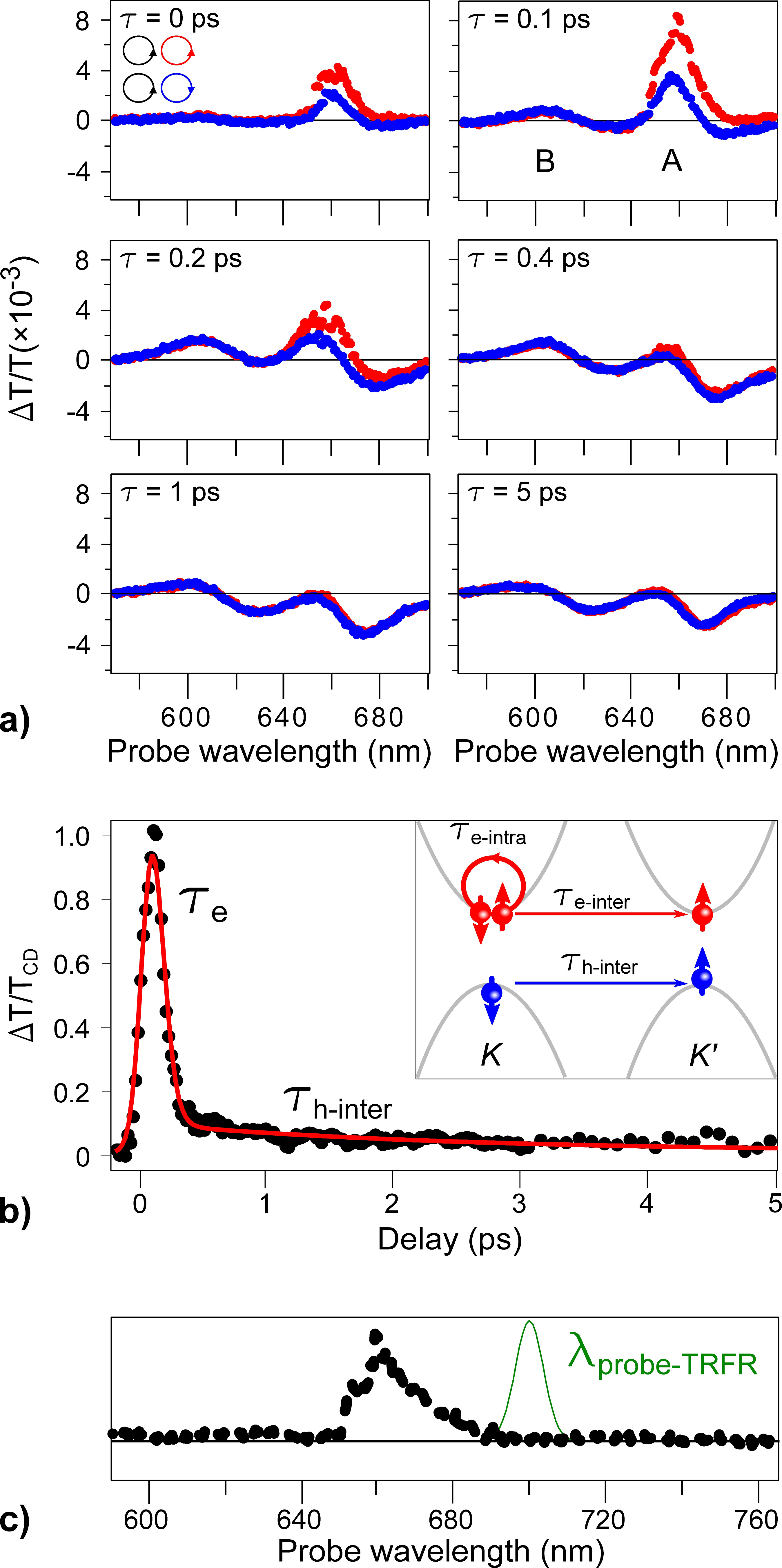}}
\caption{\label{fig2} a) $\Delta T/T$ spectra at 77K for co- and counter- circularly polarized probe pulses (red and blue traces, respectively) at different delay times. The circularly polarized narrow-band pump pulse is tuned to 650nm, while the probe pulse covers a broad energy range between 500 and 700nm. b), Black dots: difference between the $\Delta T/T$ signal at 655nm (peak of the A exciton) measured by co- and counter-circularly polarized probe pulses and normalized to unity. Red solid curve: fit with the rate equations describing the temporal dynamics of $e_\uparrow(\downarrow)(t)$ and $h_\uparrow(\downarrow)(t)$. Inset: sketch of the processes causing the time-evolution of the TRCD signal. c) Comparison between the difference between the spectral traces at $\tau=0.1$ps (black dots) and the spectrum of the probe pulse used in TRFR}
\end{figure}

In TRCD experiments the system is excited by a circularly polarized pump beam, while the transient transmittivity is probed by a co- and a counter-circularly polarized broadband probe beam. In the case of MoS$_{2}$, the pump-induced spin and valley dynamics is given by the difference of the transient signals measured in the two configurations: $\Delta T_{\mathrm{CD}}/T(\omega,t)=\Delta T_{\mathrm{SH}}/T(\omega,t)-\Delta T_{\mathrm{OH}}/T(\omega,t)$ where $\Delta T_{\mathrm{SH}(\mathrm{OH})}/T$ is the transient variation of the transmittivity measured with pump and probe pulses having the same (opposite) helicity (see Fig.\ref{fig1}b and Methods). TRCD measurements are performed at 77K to reduce the \textit{e}-\textit{h} spin scattering rate, which generally increases when room temperature is approached\cite{Lagarde2014}. The wavelength of the circularly polarized narrow-band pump pulse is tuned to 650nm (i.e. close to the direct gap optical transition measured at 77K\cite{Mai2014}) while the probe pulse covers a broad wavelength range between 500 and 700nm (see Methods). The transient optical response is dominated by the A and B excitonic transitions across \textbf{K}, centered respectively at$\sim$655 and 605nm, corresponding to the creation of an \textit{e}-\textit{h} pair with \textit{e} in the CB minimum and \textit{h} either in the VB maximum (A) or in the split-off band (B)\cite{Sim2013}. We observe a vanishing TRCD signal associated with B (Fig.\ref{fig2}c), while A shows a TRCD dynamics characterized by a bi-exponential decay (Fig.\ref{fig2}b; 655nm probe). Such behavior mirrors the different time-evolution of the spin-resolved bleaching of the direct optical transitions at \textbf{K,K'}, respectively, and can be exploited to investigate the time evolution of \textit{e}/\textit{h} spin relaxation \cite{Wang2013b,Mai2014}. In fact, in A, bleaching can occur only when the probe pulse has the same helicity as the pump, at least as long as photoexcited \textit{e}/\textit{h} maintain their original spin and valley degrees of freedom\cite{Mai2014}. The TRCD relaxation dynamics is thus determined by four processes (Fig.\ref{fig2}b): (i) \textit{h} inter-valley scattering at the VB maximum; (ii) \textit{e} inter-valley scattering and (iii) intra-valley spin flip at the CB minimum; (iv) \textit{e}-\textit{h} recombination. Note that spin-orbit coupling prohibits intra-valley \textit{h} spin flip at the VB maximum\cite{Mak2012}.

Since three relaxation channels are available to \textit{e}, while inter-valley \textit{h} scattering is a more unlikely event where momentum and spin need to be simultaneously transferred, we attribute the faster observed time scale to the \textit{e} relaxation dynamics. The temporal dynamics of the TRCD signal is analyzed in terms of the rate equations for the optically injected \textit{e}/\textit{h} spin densities, respectively $e_{\downarrow(\uparrow)}(t)$ and $h_{\uparrow(\downarrow)}(t)$.
In particular $\Delta T/T(t)$ is proportional to the product between unoccupied states in the CB $\left[n_{\mathrm{CB}}-e_{\downarrow}(t)\right]$ and occupied states in the VB $\left[n_{\mathrm{VB}}-h_{\uparrow}(t)\right]$ where $n_{\mathrm{CB}}$/$n_{\mathrm{VB}}$ are two parameters accounting for the available states in the CB/VB. The fitting function used to simulate the TRCD trace in Fig.\ref{fig2}b is: $\Delta T/T_{\mathrm{CD}}(t)\propto\{\left[n_{\mathrm{CB}}-e_{\downarrow}(t)\right]\left[n_{\mathrm{VB}}-h_{\uparrow}(t)\right]\} -\{\left[n_{\mathrm{CB}}-e_{\uparrow}(t)\right]\left[n_{\mathrm{VB}}-h_{\downarrow}(t)\right]\}$ (see Methods). It is important to stress that the above-mentioned mechanisms affecting the \textit{e}/\textit{h} dynamics are indistinguishable from the point of view of TRCD and can be described by an effective relaxation time $\tau_{e}=(1/\tau_{e-\mathrm{intra}}+1/\tau_{e-\mathrm{inter}}+1/\tau_{e-\mathrm{rec}})^{-1}\sim 100$fs for \textit{e} and $\tau_{h}=(1/\tau_{h-\mathrm{inter}}+1/\tau_{h-\mathrm{rec}})^{-1}\sim 4$ps for \textit{h}, $\tau_{e-\mathrm{intra}}$ being the time constant of the \textit{e} intra-valley (spin flip) relaxation, $\tau_{e-\mathrm{inter}}$ ($\tau_{h-\mathrm{inter}}$) the characteristic time of the inter-valley \textit{e} (\textit{h}) scattering, and $\tau_{e(h)-\mathrm{rec}}$ the \textit{e} (\textit{h}) radiative and non-radiative recombination time. The value of $\tau_{h}$ is in agreement with the spin/valley decay time ($\sim 4$~ps) measured by time-resolved PL \cite{Lagarde2014}, which sets a lower limit for the \textit{e}/\textit{h} recombination time. This reveals that recombination does not affect the \textit{e} dynamics, which is dominated by fast intra- and inter-valley scattering, and that polarized PL is given by the radiative recombination between unpolarized \textit{e} and almost fully-polarized \textit{h}.

We further investigate the inter- and intra-valley dynamics governing the non-equilibrium response of 1L-MoS$_{2}$ with TRFR, where a left (right)-circularly polarized pump pulse, resonant with the A exciton, creates a spin- and valley-polarized density of \textit{e}/\textit{h} with non-vanishing and well-defined orbital angular momentum. A delayed linearly polarized probe beam is transmitted through the sample, subsequently separated through a Wollaston prism into two orthogonal linearly-polarized beams and equally distributed on two balanced photodiodes. The transient dynamics associated with the orbital momentum of the photoexcited charges is probed by measuring the rotation angle $\theta_{F}$ of a linearly polarized probe pulse, which manifests itself as a difference between the signals measured by the two photodiodes (Fig.\ref{fig1}c). Because of the direct gap transition at the two degenerate valleys, the TRFR signal in MoS$_{2}$ is affected both by Pauli blocking, inhibiting the absorption of light having the same circular polarization of the pump, and by helicity-dependent light scattering from the photoexcited \textit{e}/\textit{h} populations. The first phenomenon is similar to that governing TRCD: \textit{e}/\textit{h} populations with well-defined spin and valley degrees of freedom reduce the absorption of probe light with the same circular polarization as the pump. Therefore, the imaginary part of the transient refractive index assumes different values for opposite light helicities. Since the real and imaginary part of the refractive index are related by the Kramers-Kronig transformations\cite{BookFox}, the system displays asymmetric values of the phase accumulated by left- and right-circularly polarized light crossing the sample, resulting in the rotation of the polarization plane of the linearly polarized probe light. This process, however, is active only when the probe light is resonant with electronic transitions across the gap. In order to rule out this effect, we perform a two-color TRFR experiment in which the probe energy is deliberately tuned below the absorption edge ($\lambda_{\mathrm{probe}}$=700nm) as shown in Fig.\ref{fig2}c.
\begin{figure}[!htbp]
\centerline{\includegraphics[width=80mm]{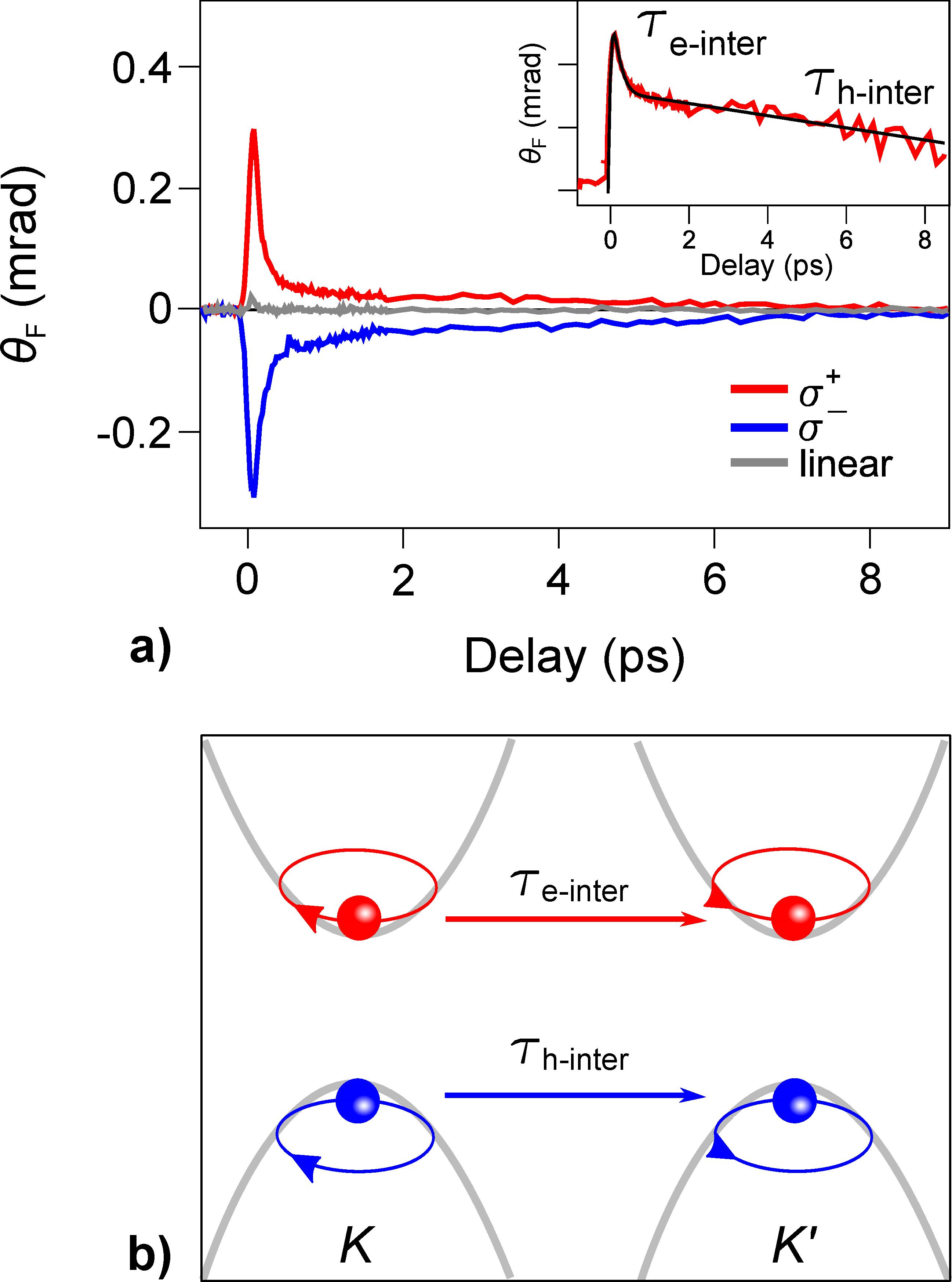}}
\caption{\label{fig3} a) TRFR traces measured by exciting the sample with left and right circularly-polarized pump pulses (red and blue curves, respectively) and a linearly polarized pump pulse (gray curve). Excitation is centered at 650nm and the probe at 700nm. All the traces are measured at 77K. The inset shows the temporal evolution of $\theta_{F}(t)$ in semi logarithmic scale in order to emphasize the bi-exponential character of the decay rate. The black solid line represents the analytical fit of the data to the rate equations. The decay constants $\tau_{e-\mathrm{inter}}$ and $\tau_{h-\mathrm{inter}}$ are related to the intervalley scattering of \textit{e} and \textit{h}, respectively. These processes are accompanied by an exchange of crystal momentum and are sketched in b).}
\end{figure}

Fig.\ref{fig3}a plots the dynamics of the Faraday rotation, $\theta_{F}(t)$, measured at 77 K by exciting spin polarized charges in either the \textbf{K} or the \textbf{K'} valley with a circularly polarized pump pulse. The change of sign of the signal with excitation helicity is due to the valley selectivity of the near gap resonant transitions. As expected, TRFR performed with a linearly polarized pump pulse (gray curve) gives rise to a negligible signal because both valleys are equally populated.

As far as electric dipole transitions are concerned, photons do not couple with the spin of the charges but only with their orbital degree of freedom \cite{Dyakonov2008}. Thus, the Faraday rotation associated with the scattering of light by the photoexcited charges can only derive from an unbalanced distribution of their orbital angular momentum projections. As shown in Fig.\ref{fig3}b, the orbital momentum of both \textit{e} and \textit{h} is associated with the valley degree of freedom, therefore the TRFR experiment is sensitive exclusively to inter-valley and recombination dynamics, and it cannot be influenced by intra-valley transitions in the CB minimum.

We now analyze the TRFR dynamics using the rate equations for the photoexcited $e_{K(K')}(t)$ and $h_{K(K')}(t)$ densities in the \textbf{K}(\textbf{K'}) valley (see Methods). The TRFR trace is then given by $\theta_{F}(t)\sim\left[A_{e}e_{K}(t)+A_{h}h_{K}(t)\right]-\left[A_{e}e_{K'}(t)+A_{h}h_{K'}(t)\right]$ where $A_{e(h)}$ are two fitting coefficients, taking into account the different weights of spin-oriented \textit{e} and \textit{h} (see Methods). As shown in the inset of Fig.\ref{fig3}a, $\theta_{F}(t)$ relaxes with two different time scales. We attribute the faster one to \textit{e} inter-valley scattering, with a decay time $\tau_{e-\mathrm{inter}}\sim 200$ fs, while we assign the slower one to \textit{h} relaxation with $\tau_{h}\sim 4$ ps, in excellent agreement with the TRCD results. With a reliable estimation for $\tau_{e-\mathrm{inter}}$ and considering that the recombination process are markedly slower than all the other \textit{e} relaxation processes (i.e. $1/\tau_{e-\mathrm{rec}}\sim 0$), we can now extrapolate from the TRCD data the characteristic time constant for intra-valley \textit{e} spin-flip transitions, obtaining $\tau_{e-\mathrm{intra}}\sim 200$ fs at 77 K, comparable to $\tau_{e-\mathrm{inter}}$.
\begin{figure*}
\centerline{\includegraphics[width=170mm]{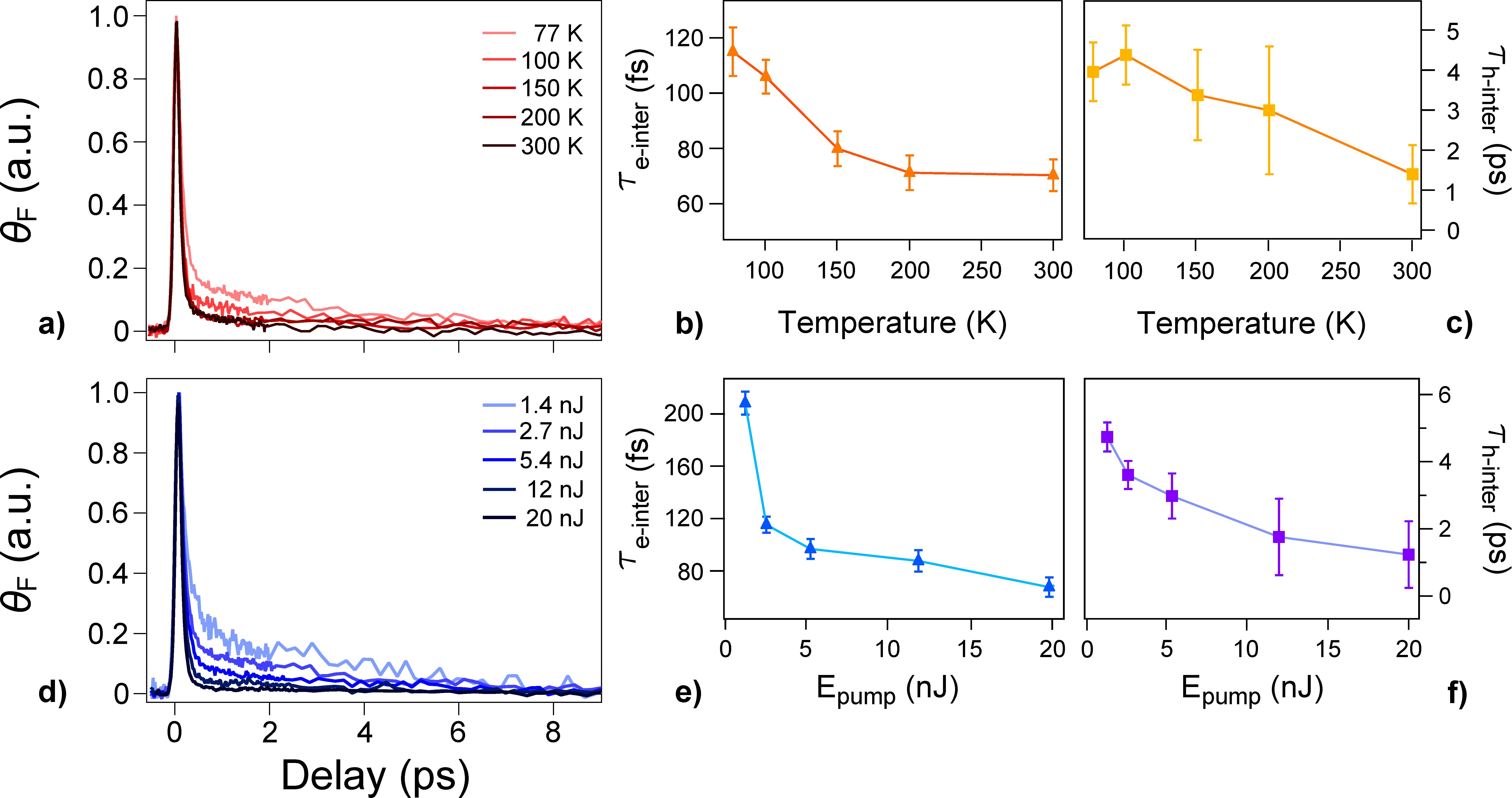}}
\caption{\label{fig4} a) Normalized TRFR dynamics measured at different temperatures in the 77-300K range. b,c) Temperature dependence of the decay constants $\tau_{e-\mathrm{inter}}$ and $\tau_{h-\mathrm{inter}}$. d), Normalized TRFR dynamics measured at different pump fluences in the 1.4-20nJ range. e,f) Fluence dependence of $\tau_{e-\mathrm{inter}}$ and $\tau_{h-\mathrm{inter}}$. $\tau_{h-\mathrm{inter}}\sim5\pm1$ ps is extrapolated from Fig.\ref{fig4}f for $E_{\mathrm{pump}}\rightarrow$0, by deconvoluting with respect to the carrier lifetime}
\end{figure*}

We then study the temperature dependence of the TRFR dynamics (Figs.\ref{fig4}a,b,c). We ascribe the observed decrease of $\tau_{e-\mathrm{inter}}$ and $\tau_{h}$ with temperature to the increase of the momentum scattering rate by phonons. The pump fluence dependence of $\theta_{F}(t)$ is also investigated in Figs.\ref{fig4}d,e,f. Both decay constants ($\tau_{e-\mathrm{inter}}$ and $\tau_{h}$) decrease exponentially for an increasing density of photoinduced charges, contrary to the case of 1L-WSe$_{2}$, where it was shown that exciton-exciton interaction does not affect the spin relaxation time\cite{Zhu2014}. We attribute this behavior to the exchange interaction acting on the \textit{e}/\textit{h} populations, which increases with pump fluence. We extrapolate the value of $\tau_{h-\mathrm{inter}}$ from $\tau_{h}$ for $E_{\mathrm{pump}}\rightarrow$ 0, obtaining $\tau_{h-\mathrm{inter},0}\sim5\pm1$ ps.

In conclusion, we determined the inter-valley and intra-valley scattering dynamics of optically injected spin-oriented charges in 1L-MoS$_{2}$ by combining two complementary techniques, TRCD and TRFR. We found that the characteristic time scales of spin-flip and inter-valley scattering for electrons are both$\sim200$fs, whereas hole inter-valley scattering acts on a time scale$\sim5$ps. The large stabilization of spin/valley degrees of freedom resulting from the coupling between spin and valley dynamics paves the way to the integration of materials with large spin-orbit in robust spintronic/valleytronic platforms. Furthermore, the procedure described here is of general validity and can be applied to study \textit{e}/\textit{h} intra and intervalley scattering processes in any other monolayer TMD.

\begin{acknowledgments}
We acknowledge funding from Nanofacility Piemonte, ERC Synergy Hetero2D, Graphene Flagship (n. 604391), a Royal Society Wolfson Research Merit Award, EPSRC grants EP/K01711X/1, EP/K017144/1, EP/L016087/1, Grant N. 2013–0615 (Fondazione Cariplo), Futuro in Ricerca grant No. RBFR12SW0J of the Italian Ministry of Education, University and Research and SEARCH-IV No. 2013-0623, grant of the Fondazione Cariplo.
\end{acknowledgments}

\section{Methods}
\setcounter{figure}{0}
\setcounter{equation}{0}
\makeatletter
\renewcommand{\thefigure}{M\@arabic\c@figure}
\makeatother
\makeatletter
\renewcommand{\theequation}{M\@arabic\c@equation}
\makeatother

\subsection{Sample Preparation}
\begin{figure}
\centerline{\includegraphics[width=40mm]{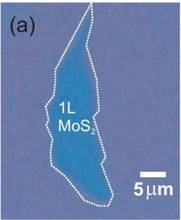}}
\centerline{\includegraphics[width=60mm]{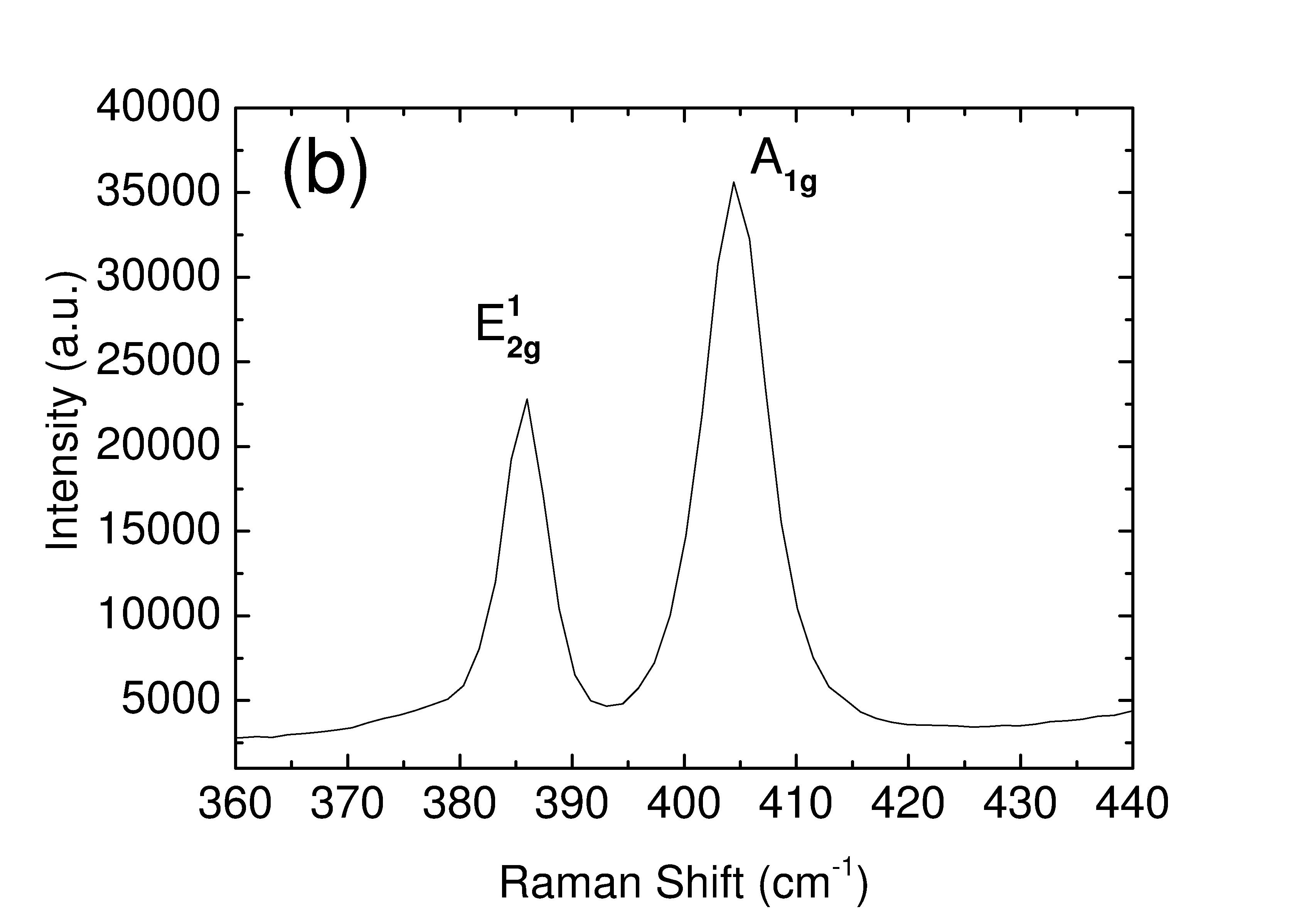}}
\centerline{\includegraphics[width=60mm]{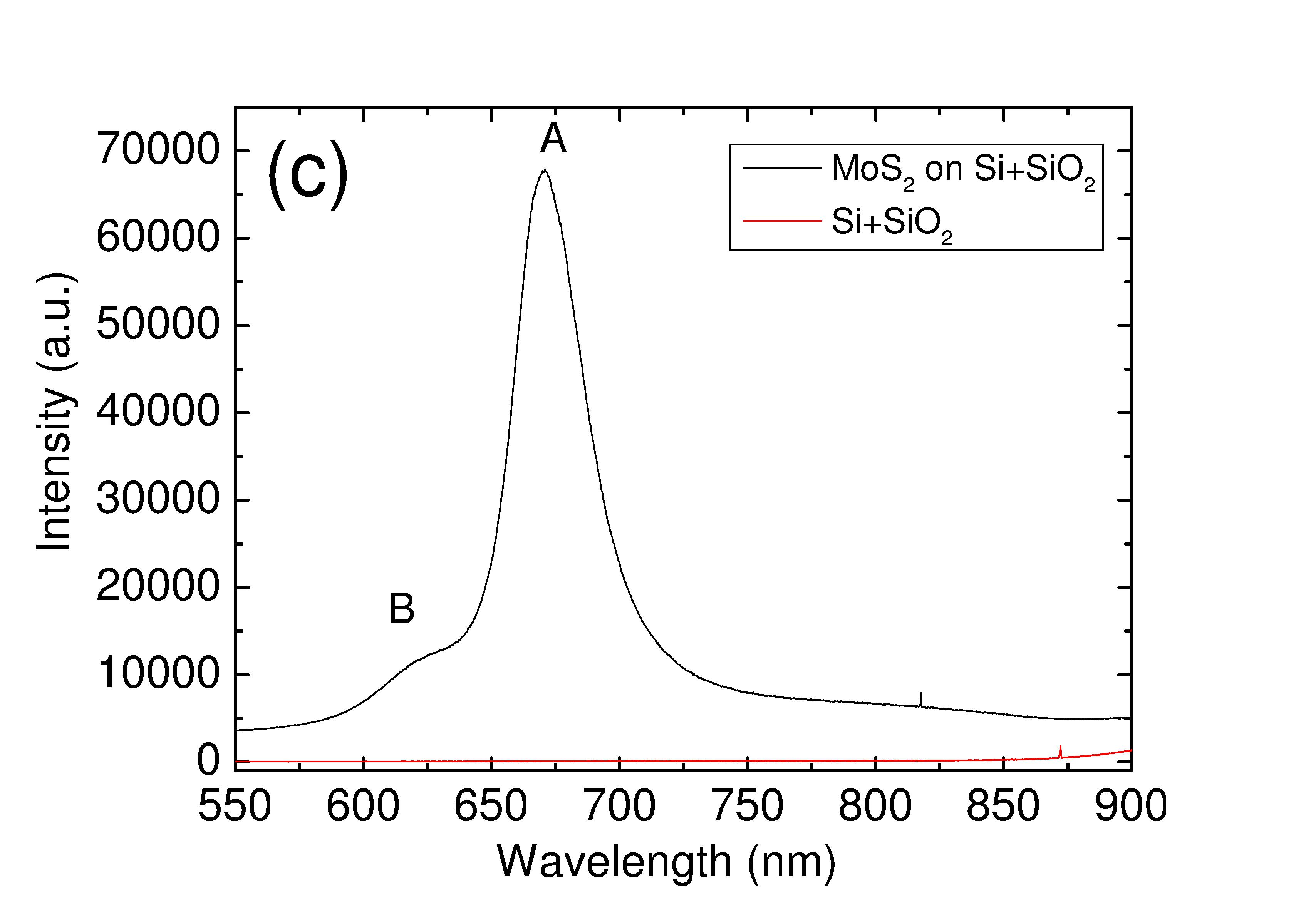}}
\caption{\label{figS1}a) Selected 1L-MoS$_{2}$ flake on SiO$_{2}$. b) Raman spectrum of the flake used for our experiments for 514.5nm excitation. c) PL spectrum of the selected flake measured for 514.5 nm excitation measured at room temperature.
}
\end{figure}

MoS$_{2}$ flakes are produced by micromechanical cleavage of bulk MoS$_{2}$ crystals (Structure Probe Inc.-SPI, natural molybdenite) onto Si wafers covered with 285nm SiO$_{2}$. 1L-MoS$_{2}$ flakes (Fig.\ref{figS1}a) are then identified by optical contrast, PL and Raman spectroscopy\cite{Lee2010,Zhang2013}. Fig.\ref{figS1}b shows a representative Raman spectrum of the flake used for our experiment, acquired at 514.5nm with a Renishaw microspectrometer. The position of the two main Raman peaks is$\sim385$ and$\sim404$cm$^{-1}$, indicating that this is 1L-MoS$_{2}$\cite{Lee2010,Zhang2013,Sundaram2013}. Fig.\ref{figS1}c plots the PL spectrum of our flake measured at room temperature for 514.5nm excitation. This consists of two bands at$\sim 1.85$ and$\sim 1.98$eV, consistent with the A and B excitons in 1L-MoS$_{2}$\cite{Splendiani2010}.
The selected 1L-MoS$_{2}$ flake is then moved by a wet-transfer technique based on a sacrificial layer of poly-methyl-methacrylate (PMMA)\cite{Bonaccorso2012,Scheneider2010}. The polymer is deposited onto the flake by spin coating, followed by immersion in de-ionized water. Water intercalation at the PMMA-SiO$_{2}$ interface detaches the polymer film\cite{Bonaccorso2012,Scheneider2010}, with the 1L-MoS$_{2}$ flake attached on it. This is then moved onto a 100 $\mu$m thick fused silica substrate and left to dry for a few hours. The fused silica substrate is particularly suitable for TRFR experiments since it is an isotropic material, without crystal orientation, and it does not exhibit birefringence\cite{ref_index}. PMMA is then removed by acetone, and the flake is released onto the fused silica\cite{Bonaccorso2012,Scheneider2010}.
A metal frame is then fabricated around the selected 1L-MoS$_{2}$ by photolithography, followed by thermal evaporation of 2nm Cr and 100nm Au films. This ensures the same flake can be easily found for TRCD and TRFR measurements. To ascertain that no damage or changes are induced on the selected flake by the transfer process, this is further characterized after transfer on fused silica and after the metal frame is defined, but before TRCD and TRFR. Fig.\ref{figS1} shows the selected 1L-MoS$_{2}$ transferred on glass and its Raman and PL spectra. These show no significant changes with respect to the initial sample prior to transfer.
\begin{figure}
\centerline{\includegraphics[width=55mm]{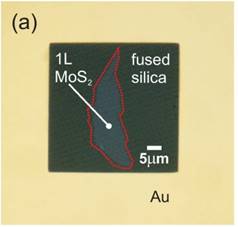}}
\centerline{\includegraphics[width=60mm]{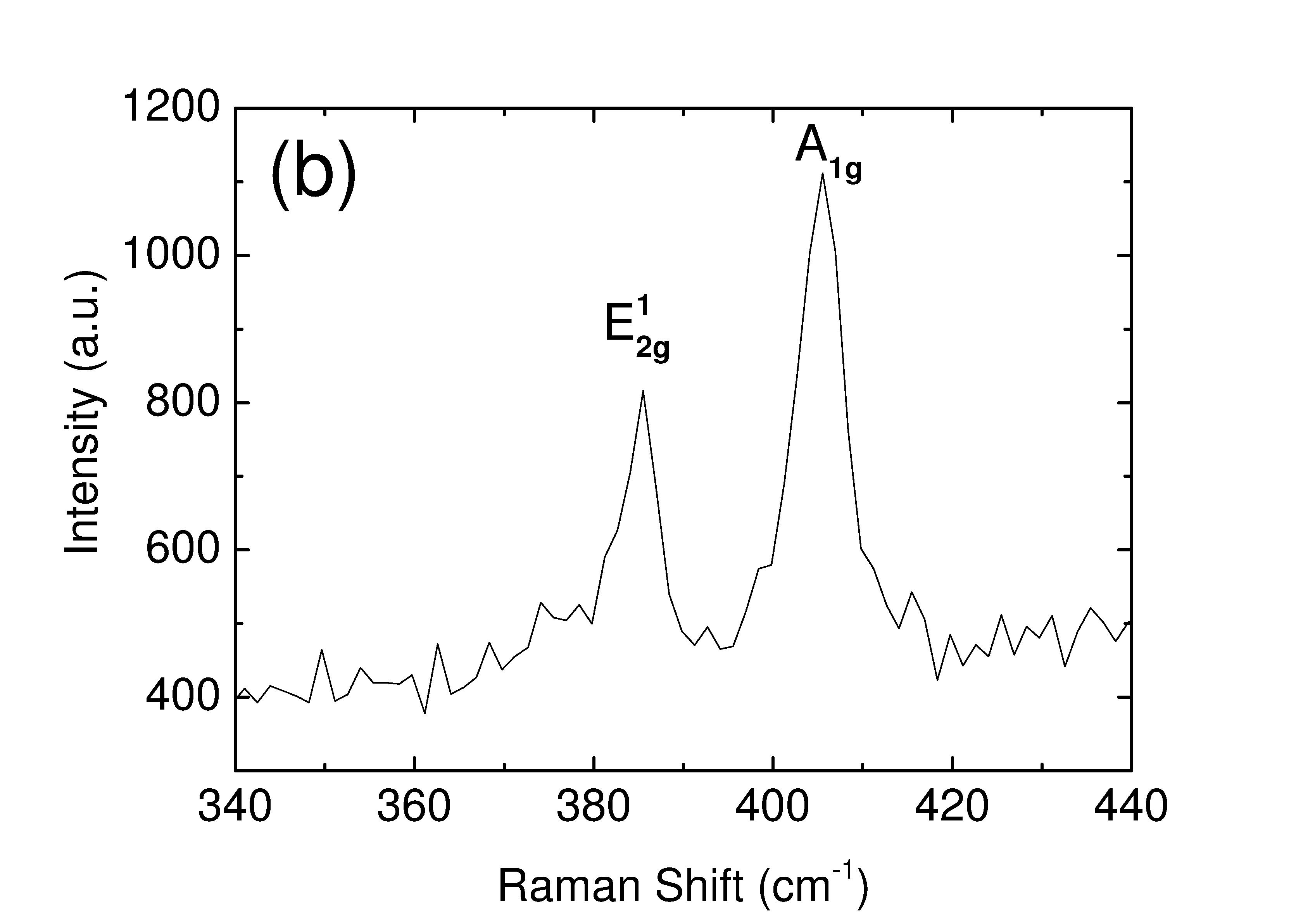}}
\centerline{\includegraphics[width=60mm]{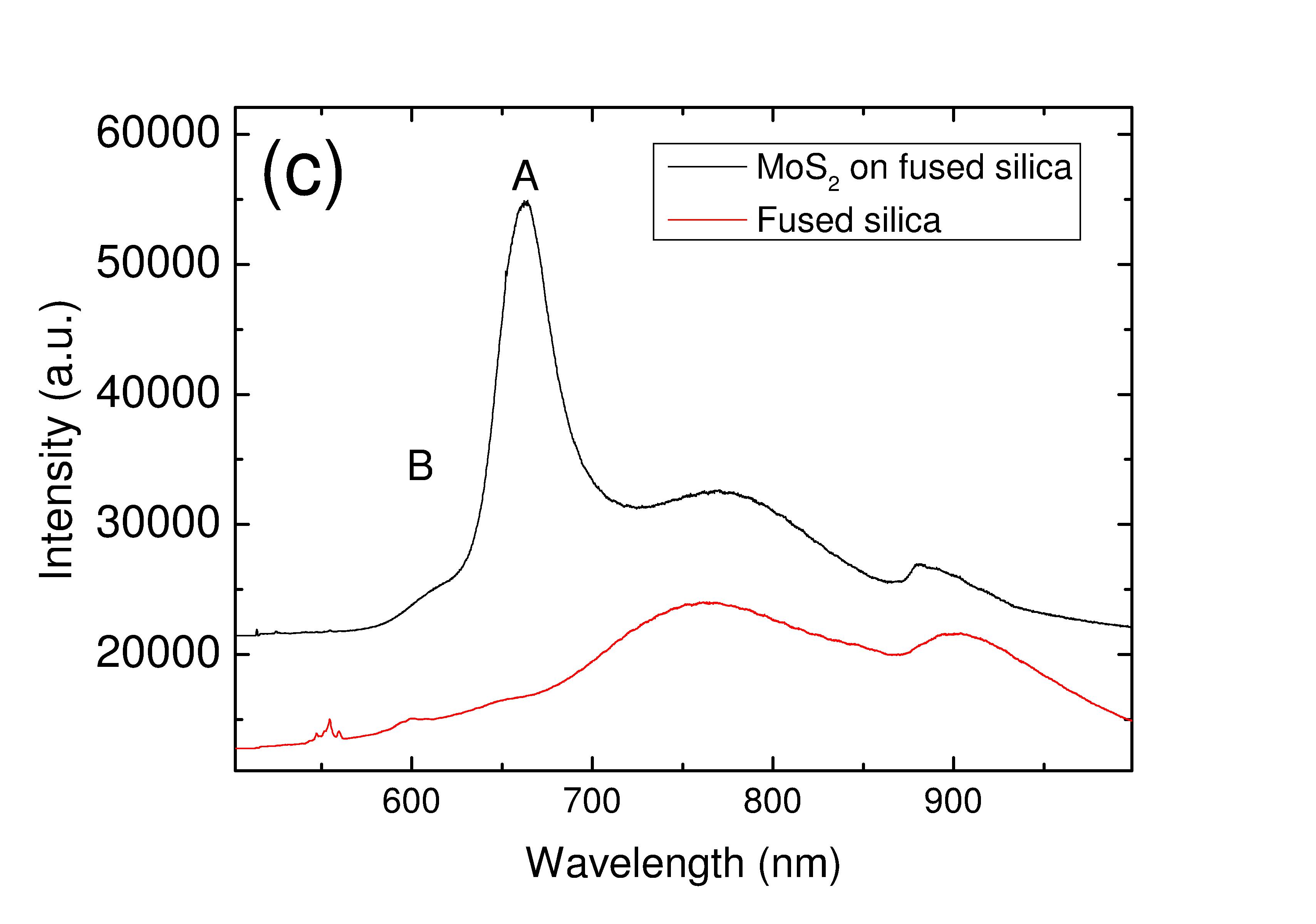}}
\caption{\label{figS2} a) Selected 1L-MoS$_{2}$ flake after transfer on fused silica. b) Raman spectrum of the flake in (a) measured for 514.5 nm excitation. c) PL spectrum of the flake in (a) for 514.5 nm excitation measured at room temperature.
}
\end{figure}
\subsection{Time-resolved Faraday rotation}
A regeneratively amplified mode locked Ti:Sapphire laser, providing 150 fs, 500$\mu$J pulses at 780nm and 1kHz repetition rate drives two optical parametric amplifiers (OPA). The output of the first OPA, used to inject spin polarized carriers in the sample, is tuned to 650nm, i.e. quasi resonant to the bandgap wavelength, and it is circularly polarized by a quarter wave plate. The probe pulse, provided by the second OPA, is linearly polarized and it is centered at 700nm. Both pump and probe pulses have a spectral bandwidth of 10 nm. The temporal delay between the two pulses is adjusted by a mechanical delay stage. At the sample position the pump(probe) diameter is 50(20)$\mu$m. We estimate a temporal resolution $\Gamma_{\mathrm{TRFR}}$ of the TRFR setup$\sim70$fs. The sample is positioned in a liquid nitrogen cryostat and the temperature is checked by a thermocouple placed in proximity to the sample. The transmitted probe pulse passes through a Wollaston prism and it is focused on a couple of balanced photodiodes. The Wollaston prism is rotated in order to equalize the probe intensities on the two photodiodes. The pump induced imbalance of the signal measured by the two photodiodes is registered by a lock-in amplifier which is locked to the modulation frequency of the pump beam (i.e. 500Hz).
\subsection{Time-resolved circular dichroism}
The pump beam is generated by an OPA at 650nm and it is circularly polarized. The probe is a white light pulse, ranging between 500 and 730nm, generated by focusing the fundamental pulse in a 2mm-thick sapphire plate. The probe pulse is circularly polarized by a broadband quarter wave plate. We estimate the temporal resolution of the TRCD setup to be$\sim200$fs for the entire spectral window of the probe beam. A fast analog-to-digital conversion card with 16-bit resolution enables single-shot recording of the probe spectrum at the full 1kHz repetition rate. By recording the transmitted spectra with ($T_{\mathrm{on}}$) and without ($T_{\mathrm{off}}$) the pump excitation at different delay times one obtains a two-dimensional differential transmission ($\Delta T/T$) map as as a function of probe frequency $\omega$ and delay $\tau$:
\begin{equation*}\label{pumpprobe}
\Delta T(\omega,\tau)=\frac{T_{\mathrm{on}}(\omega,\tau)-T_{\mathrm{off}}(\omega)}{T_{\mathrm{off}}(\omega)}.
\end{equation*}
\subsection{Rate equations for spin oriented electron and hole densities}
TRCD results can be interpreted in terms of the time evolution of the spin oriented \textit{e} and \textit{h} densities, which are optically injected in 1L-MoS$_{2}$ through optical orientation. Let $e_{\uparrow(\downarrow)}(t)$ be the \textit{e} densities generated at \textbf{K}, \textbf{K'} and $h_{\downarrow(\uparrow)}(t)$ the \textit{h} densities at \textbf{K}, \textbf{K'}. We describe the temporal evolution of the spin polarized \textit{e}/\textit{h} densities, following an excitation with left-circularly polarized light, with a set of four coupled rate equations, similar to those used to model the MoS$_{2}$ exciton relaxation rate in Refs.\onlinecite{Mak2012,Kioseoglou2012}:
\begin{gather}
\frac{de_{\downarrow}(t)}{dt}=S\left(t\right)-\left(e_{\downarrow}(t)-e_{\uparrow}(t)\right)\left(\frac{1}{\tau_{e-\mathrm{intra}}} +\frac{1}{\tau_{e-\mathrm{inter}}}\right)\\
\frac{de_{\uparrow}(t)}{dt}=\left(e_{\downarrow}(t)-e_{\uparrow}(t)\right)\left(\frac{1}{\tau_{e-\mathrm{intra}}}+ \frac{1}{\tau_{e-\mathrm{inter}}}\right)\\
\frac{dh_{\uparrow}(t)}{dt}=S\left(t\right)-\frac{h_{\uparrow}(t)-h_{\downarrow}(t)}{\tau_{h-\mathrm{inter}}}- \frac{h_{\uparrow}(t)}{\tau_{h-\mathrm{rec}}}\\
\frac{dh_{\downarrow}(t)}{dt}=\frac{h_{\uparrow}(t)-h_{\downarrow}(t)}{\tau_{h-\mathrm{inter}}}- \frac{h_{\downarrow}(t)}{\tau_{h-\mathrm{rec}}}.
\end{gather}
The gaussian function $S\left(t\right)$ represents the laser source term which accounts for the temporal resolution of the TRCD setup. The transient depletion of the $e_{\downarrow}$ and $h_{\uparrow}$ carriers (Eqs.M1, M3), injected by the pump in a valley, and the consequent filling of the reversed spin polarized carriers $e_{\uparrow}$ and $h_{\downarrow}$ (Eqs.M2, M4) are regulated by spin-flip processes. For \textit{e} these processes can occur either in the same valley, or by scattering in the opposite valley. The former scattering process is regulated by the time constant $\tau_{e-\mathrm{intra}}$, while the latter depends also on $\tau_{e-\mathrm{inter}}$. For \textit{h} spin and valley relaxation coincide, and spin reversal processes are accompanied by an exchange of linear momentum ($\tau_{h-\mathrm{inter}}$). 1$/\tau_{h-\mathrm{rec}}$ is the \textit{h} recombination rate, while the \textit{e} recombination rate is one order of magnitude slower than the intra- and inter-valley scattering rates, thus it can be safely neglected in Eqs.M1, M2 (1$/\tau_{e-\mathrm{rec}}\sim 0$). Once the Eqs.M1-M4 are solved, the TRCD curves are generated by the product of the unoccupied states at the top of the valence band and unoccupied states at the bottom of the conduction band:
\begin{multline}\label{TRCDeq}
\left[\frac{\Delta T(t)}{T}\right]_{\mathrm{CD}}\sim\left\{\left[n_{\mathrm{CB}}-e_{\downarrow}(t)\right]\left[n_{\mathrm{VB}}-h_{\uparrow}(t)\right]\right\}+\\
-\left\{\left[n_{\mathrm{CB}}-e_{\uparrow}(t)\right]\left[n_{\mathrm{VB}}-h_{\downarrow}(t)\right]\right\},
\end{multline}
where $n_{\mathrm{CB}}$ and $n_{\mathrm{VB}}$ are free fitting parameters accounting for the available states in CB and VB.

Considering TRFR, the set of four rate equations has to be slightly modified in order to take into account the sensitivity of the measurement with respect to the photogenerated densities of \textit{e}/\textit{h} in \textbf{K} and \textbf{K'}. Let $e_{K(K')}(t)$ be the \textit{e} densities generated at \textbf{K}, \textbf{K'} and $h_{K(K')}(t)$ the \textit{h} densities at \textbf{K}, \textbf{K'}. In this case, considering an excitation with left-circularly polarized light, the set of four coupled rate equations can be written as:
\begin{gather}
\frac{de_{K}(t)}{dt}=S\left(t\right)-\frac{e_{K}(t)-e_{K'}(t)}{\tau_{e-\mathrm{inter}}}\\
\frac{de_{K'}(t)}{dt}=\frac{e_{K}(t)-e_{K'}(t)}{\tau_{e-\mathrm{inter}}}\\
\frac{dh_{K}(t)}{dt}=S\left(t\right)-\frac{h_{K}(t)-h_{K'}(t)}{\tau_{h-\mathrm{inter}}}-\frac{h_{K}(t)}{\tau_{h-\mathrm{rec}}}\\
\frac{dh_{K'}(t)}{dt}=\frac{h_{K}(t)-h_{K'}(t)}{\tau_{h-\mathrm{inter}}}-\frac{h_{K'}(t)}{\tau_{h-\mathrm{rec}}},
\end{gather}
where the full-width half-maximum $\Gamma_{\mathrm{TRFR}}$ of the gaussian function takes into account the temporal resolution of the TRFR setup. The TRFR is only sensitive to the \textit{e}/\textit{h} intervalley scattering time, which affects the unbalance of orbital angular momentum, optically injected by circularly polarized light so that the TRFR fit function can be written as:
\begin{equation}\label{TRFReq}
\theta_{F}(t)\sim \left[A_{e}e_{K}(t)+A_{h}h_{K}(t)\right]-\left[A_{e}e_{K'}(t)+A_{h}h_{K'}(t)\right],
\end{equation}
where A$_{e(h)}$ are two free fitting parameters whose ratio is varied in order to obtain the best fit of the data.
The simultaneous fit of the TRCD and TRFR traces with Eqs.M5, M10 enables us to determine the time constants for intra- and inter-valley scattering for both electrons and holes.

\end{document}